
%
\message
{TCMJNL macro package version 1.0 as of 1/2/88. If there is a severe bug,
consult Mike Cates.}
\catcode`@=11
\expandafter\ifx\csname inp@t\endcsname\relax\let\inp@t=\input
\def\input#1 {\expandafter\ifx\csname #1IsLoaded\endcsname\relax
\inp@t#1%
\expandafter\def\csname #1IsLoaded\endcsname{(#1 was previously loaded)}
\else\message{\csname #1IsLoaded\endcsname}\fi}\fi
\catcode`@=12



\font\twelverm=cmr10 scaled 1200    \font\twelvei=cmmi10 scaled 1200
\font\twelvesy=cmsy10 scaled 1200   \font\twelveex=cmex10 scaled 1200
\font\twelvebf=cmbx10 scaled 1200   \font\twelvesl=cmsl10 scaled 1200
\font\twelvett=cmtt10 scaled 1200   \font\twelveit=cmti10 scaled 1200
\font\twelvesc=cmbx10 scaled 1200  \font\twelvesf=cmbx10 scaled 1200
\skewchar\twelvei='177   \skewchar\twelvesy='60


\def\twelvepoint{\normalbaselineskip=12.4pt plus 0.1pt minus 0.1pt
  \abovedisplayskip 12.4pt plus 3pt minus 9pt
  \belowdisplayskip 12.4pt plus 3pt minus 9pt
  \abovedisplayshortskip 0pt plus 3pt
  \belowdisplayshortskip 7.2pt plus 3pt minus 4pt
  \smallskipamount=3.6pt plus1.2pt minus1.2pt
  \medskipamount=7.2pt plus2.4pt minus2.4pt
  \bigskipamount=14.4pt plus4.8pt minus4.8pt
  \def\rm{\fam0\twelverm}          \def\it{\fam\itfam\twelveit}%
  \def\sl{\fam\slfam\twelvesl}     \def\bf{\fam\bffam\twelvebf}%
  \def\mit{\fam 1}                 \def\cal{\fam 2}%
  \def\sc{\twelvesc}		   \def\tt{\twelvett}
  \def\sf{\twelvesf}
  \textfont0=\twelverm   \scriptfont0=\tenrm   \scriptscriptfont0=\sevenrm
  \textfont1=\twelvei    \scriptfont1=\teni    \scriptscriptfont1=\seveni
  \textfont2=\twelvesy   \scriptfont2=\tensy   \scriptscriptfont2=\sevensy
  \textfont3=\twelveex   \scriptfont3=\twelveex  \scriptscriptfont3=\twelveex
  \textfont\itfam=\twelveit
  \textfont\slfam=\twelvesl
  \textfont\bffam=\twelvebf \scriptfont\bffam=\tenbf
  \scriptscriptfont\bffam=\sevenbf
  \normalbaselines\rm}



\def\beginlinemode{\endmode
  \begingroup\parskip=0pt \obeylines\def\\{\par}\def\endmode{\par\endgroup}}
\def\beginparmode{\endmode
  \begingroup \def\endmode{\par\endgroup}}
\let\endmode=\par
{\obeylines\gdef\
{}}
\def\singlespace{\baselineskip=\normalbaselineskip}

\def\oneandahalfspace{\baselineskip=\normalbaselineskip
  \multiply\baselineskip by 3 \divide\baselineskip by 2}
\def\doublespace{\baselineskip=\normalbaselineskip \multiply\baselineskip by 2}

\newcount\firstpageno
\firstpageno=2
\footline={\ifnum\pageno<\firstpageno{\hfil}\else{\hfil\twelverm\folio\hfil}\fi}
\def\toppageno{\global\footline={\hfil}\global\headline
  ={\ifnum\pageno<\firstpageno{\hfil}\else{\hfil\twelverm\folio\hfil}\fi}}
\let\rawfootnote=\footnote		
\def\footnote#1#2{{\rm\singlespace\parindent=0pt\parskip=0pt
  \rawfootnote{#1}{#2\hfill\vrule height 0pt depth 6pt width 0pt}}}
\def\raggedcenter{\leftskip=4em plus 12em \rightskip=\leftskip
  \parindent=0pt \parfillskip=0pt \spaceskip=.3333em \xspaceskip=.5em
  \pretolerance=9999 \tolerance=9999
  \hyphenpenalty=9999 \exhyphenpenalty=9999 }
\def\dateline{\rightline{\ifcase\month\or
  January\or February\or March\or April\or May\or June\or
  July\or August\or September\or October\or November\or December\fi
  \space\number\year}}
\def\received{\vskip 3pt plus 0.2fill
 \centerline{\sl (Received\space\ifcase\month\or
  January\or February\or March\or April\or May\or June\or
  July\or August\or September\or October\or November\or December\fi
  \qquad, \number\year)}}


\hsize=6.2truein
\hoffset=0.0truein
\parskip=\medskipamount
\def\\{\cr}
\twelvepoint		
\oneandahalfspace		
\overfullrule=0pt	




\def\title			
  {\null\vskip 3pt plus 0.2fill
   \beginlinemode \doublespace \raggedcenter \bf}

\def\author			
  {\vskip 13pt plus 0.1fill \beginlinemode
   \singlespace \raggedcenter\rm}

\def\affil			
  {\vskip 3pt plus 0.1fill \beginlinemode
   \oneandahalfspace \raggedcenter \sl}

\def\abstract			
  {\vskip 3pt plus 0.3fill \beginparmode
   \oneandahalfspace ABSTRACT: }

\def\endtitlepage		
  {\endpage			
   \body}
\let\endtopmatter=\endtitlepage

\def\body			
  {\beginparmode}		

\def\head#1{			
  \goodbreak\vskip 0.5truein	
  {\immediate\write16{#1}
   \raggedcenter \uppercase{#1}\par}
   \nobreak\vskip 0.25truein\nobreak}

\def\beginitems{
\par\medskip\bgroup\def\i##1 {\item{##1}}\def\ii##1 {\itemitem{##1}}
\leftskip=36pt\parskip=0pt}
\def\enditems{\par\egroup}

\def\beneathrel#1\under#2{\mathrel{\mathop{#2}\limits_{#1}}}

\def\refto#1{$^{#1}$}		

\def\references			
  {\head{References}		
   \beginparmode
   \frenchspacing \parindent=0pt \leftskip=1truecm
   \parskip=8pt plus 3pt \everypar{\hangindent=\parindent}}

\gdef\refis#1{\item{#1.\ }}			

\gdef\journal#1, #2, #3, 1#4#5#6{		
    {\sl #1~}{\bf #2}, #3 (1#4#5#6)}		

\def\endreferences{\body}

\def\figurecaptions		
  {\endpage
   \beginparmode
   \head{Figure Captions}
}

\def\endpage			
  {\vfill\eject}

\def\endpaper			
  {\endmode\vfill\supereject}


\def\heading				
  {\vskip 0.5truein plus 0.1truein	
   \beginparmode \def\\{\par} \parskip=0pt \singlespace \raggedcenter}

\def\subheading				
  {\vskip 0.25truein plus 0.1truein	
   \beginlinemode \singlespace \parskip=0pt \def\\{\par}\raggedcenter}

\def\tag#1$${\eqno(#1)$$}

\def\align#1$${\eqalign{#1}$$}

\def\aligntag#1$${\gdef\tag##1\\{&(##1)\cr}\eqalignno{#1\\}$$
  \gdef\tag##1$${\eqno(##1)$$}}

\def\overset #1\to#2{{\mathop{#2}\limits^{#1}}}
\def\underset#1\to#2{{\let\next=#1\mathpalette\undersetpalette#2}}
\def\undersetpalette#1#2{\vtop{\baselineskip0pt
\ialign{$\mathsurround=0pt #1\hfil##\hfil$\crcr#2\crcr\next\crcr}}}


\def\ref#1{Ref.~#1}			
\def\Ref#1{Ref.~#1}			
\def\[#1]{[\cite{#1}]}
\def\cite#1{{#1}}
\def\(#1){(\call{#1})}
\def\call#1{{#1}}
\def\taghead#1{}
\def\frac#1#2{{#1 \over #2}}

\def\12{{1\over2}}

\def\sla{\raise.15ex\hbox{$/$}\kern-.57em}
\def\leaderfill{\leaders\hbox to 1em{\hss.\hss}\hfill}
\def\twiddle{\lower.9ex\rlap{$\kern-.1em\scriptstyle\sim$}}
\def\bigtwiddle{\lower1.ex\rlap{$\sim$}}
\def\gtwid{\mathrel{\raise.3ex\hbox{$>$\kern-.75em\lower1ex\hbox{$\sim$}}}}
\def\ltwid{\mathrel{\raise.3ex\hbox{$<$\kern-.75em\lower1ex\hbox{$\sim$}}}}
\def\square{\kern1pt\vbox{\hrule height 1.2pt\hbox{\vrule width 1.2pt\hskip 3pt
   \vbox{\vskip 6pt}\hskip 3pt\vrule width 0.6pt}\hrule height 0.6pt}\kern1pt}
\def\tdot#1{\mathord{\mathop{#1}\limits^{\kern2pt\ldots}}}

\def\pmb#1{\setbox0=\hbox{#1}%
  \kern-.025em\copy0\kern-\wd0
  \kern  .05em\copy0\kern-\wd0
  \kern-.025em\raise.0433em\box0 }

\catcode`@=11
\newcount\r@fcount \r@fcount=0
\newcount\r@fcurr
\immediate\newwrite\reffile
\newif\ifr@ffile\r@ffilefalse
\def\w@rnwrite#1{\ifr@ffile\immediate\write\reffile{#1}\fi\message{#1}}

\def\writer@f#1>>{}
\def\referencefile{
  \r@ffiletrue\immediate\openout\reffile=\jobname.ref%
  \def\writer@f##1>>{\ifr@ffile\immediate\write\reffile%
    {\noexpand\refis{##1} = \csname r@fnum##1\endcsname = %
     \expandafter\expandafter\expandafter\strip@t\expandafter%
     \meaning\csname r@ftext\csname r@fnum##1\endcsname\endcsname}\fi}%
  \def\strip@t##1>>{}}

\def\citeall#1{\xdef#1##1{#1{\noexpand\cite{##1}}}}
\def\cite#1{\each@rg\citer@nge{#1}}	

\def\each@rg#1#2{{\let\thecsname=#1\expandafter\first@rg#2,\end,}}
\def\first@rg#1,{\thecsname{#1}\apply@rg}	
\def\apply@rg#1,{\ifx\end#1\let\next=\relax
\else,\thecsname{#1}\let\next=\apply@rg\fi\next}

\def\citer@nge#1{\citedor@nge#1-\end-}	
\def\citer@ngeat#1\end-{#1}
\def\citedor@nge#1-#2-{\ifx\end#2\r@featspace#1 
  \else\citel@@p{#1}{#2}\citer@ngeat\fi}	
\def\citel@@p#1#2{\ifnum#1>#2{\errmessage{Reference range #1-#2\space is bad.}%
    \errhelp{If you cite a series of references by the notation M-N, then M and
    N must be integers, and N must be greater than or equal to M.}}\else%
 {\count0=#1\count1=#2\advance\count1
by1\relax\expandafter\r@fcite\the\count0,%
  \loop\advance\count0 by1\relax
    \ifnum\count0<\count1,\expandafter\r@fcite\the\count0,%
  \repeat}\fi}

\def\r@featspace#1#2 {\r@fcite#1#2,}	
\def\r@fcite#1,{\ifuncit@d{#1}
    \newr@f{#1}%
    \expandafter\gdef\csname r@ftext\number\r@fcount\endcsname%
                     {\message{Reference #1 to be supplied.}%
                      \writer@f#1>>#1 to be supplied.\par}%
 \fi%
 \csname r@fnum#1\endcsname}
\def\ifuncit@d#1{\expandafter\ifx\csname r@fnum#1\endcsname\relax}%
\def\newr@f#1{\global\advance\r@fcount by1%
    \expandafter\xdef\csname r@fnum#1\endcsname{\number\r@fcount}}

\let\r@fis=\refis			
\def\refis#1#2#3\par{\ifuncit@d{#1}
   \newr@f{#1}%
   \w@rnwrite{Reference #1=\number\r@fcount\space is not cited up to now.}\fi%
  \expandafter\gdef\csname r@ftext\csname r@fnum#1\endcsname\endcsname%
  {\writer@f#1>>#2#3\par}}

\def\ignoreuncited{
   \def\refis##1##2##3\par{\ifuncit@d{##1}%
     \else\expandafter\gdef\csname r@ftext\csname
r@fnum##1\endcsname\endcsname%
     {\writer@f##1>>##2##3\par}\fi}}

\def\r@ferr{\endreferences\errmessage{I was expecting to see
\noexpand\endreferences before now;  I have inserted it here.}}
\let\r@ferences=\references
\def\references{\r@ferences\def\endmode{\r@ferr\par\endgroup}}

\let\endr@ferences=\endreferences
\def\endreferences{\r@fcurr=0
  {\loop\ifnum\r@fcurr<\r@fcount
    \advance\r@fcurr by 1\relax\expandafter\r@fis\expandafter{\number\r@fcurr}%
    \csname r@ftext\number\r@fcurr\endcsname%
  \repeat}\gdef\r@ferr{}\endr@ferences}


\let\r@fend=\endpaper\gdef\endpaper{\ifr@ffile
\immediate\write16{Cross References written on []\jobname.REF.}\fi\r@fend}

\catcode`@=12

\citeall\refto		
\citeall\ref		%
\citeall\Ref		%


\title{ \bf
SELF INDUCED QUENCHED DISORDER: A MODEL FOR THE GLASS TRANSITION
 }

\author{J.P. Bouchaud }
\affil{Service
de Physique de l'Etat Condens\'e, CEA-Saclay, Orme des Merisiers, 91 191 Gif
s/ Yvette CEDEX}
\author{M. M\'ezard }
\affil{Laboratoire de Physique Th\'eorique, Ecole Normale Sup\'erieure, 24 rue
Lhomond 75231 Paris CEDEX 05 (Unit\'e propre du CNRS, associ\'ee
\`a l'ENS et \`a l'Universit\'e de Paris sud) }

\abstract {
We consider a simple spin system without disorder
 which exhibits a glassy regime.
We show that this model can be well approximated by a system with quenched
disorder which is studied with the standard methods developped in spin glasses.
We propose that the glass transition is a point where quenched disorder is self
induced,
 a scenario for which the `cavity' method might be particularly well suited.

}

\endtopmatter
\par
The problem of the glass state remains one  major unsolved issues in condensed
matter theory. Despite an enormous body of experimental and numerical data and
quite
detailed phenomenological theories [\cite{Houches},\cite{AHV},\cite{Thiru}],
 there is no fully
satisfactory microscopic model for the glass state. The intense theoretical
activity on
spin-glasses and other disordered systems [\cite{MPV}] stemed in part because
they
retain `half' of the complexity of glasses: given a disordered (`quenched') set
of
interactions, what is the thermodynamics of the `spin' degrees of freedom, is
there a
low temperature spin glass phase, etc... The spin glass theory has indeed given
birth
to many seminal ideas which have been transfered to other glassy systems like
proteins
[\cite{Orland},\cite{Shak},\cite{Wolynes}] , rubber [\cite{Goldbart}], or even
glass
itself [\cite{Thiru}]. One subtle aspect of glasses is
that there is no  clear a priori distinction between
`slow' degrees of freedom responsible for
random interactions and `fast' degrees of freedom equilibrating therein,
although
everything goes as if it was the case: quenched disorder is self-induced. A
satisfactory glass theory requires a detailed  mathematical description of this
scenario (in fact implicit in
the mode coupling theory [\cite{Gotze}]) and the identification of these `slow'
degrees of freedom .

In the following we shall show a possible way to get round this
problem.
 We present a simple model, already
studied in [\cite{Golay},\cite{Berna}], which contains no quenched disorder.
At this stage, the only justification for
studying this model comes from numerical simulations : the system very clearly
exhibits features of a glass transition
 - jump in the specific heat [\cite{Berna}],
slow dynamics and aging [\cite{MW}], etc... We propose an unusual analytical
approach
to this model which is to find a `fiduciary' disordered model which is `as
close as
possible' to the pure model, but for which all the ideas and methods developed
for
spin glasses (replicas, cavity method, statistics of the metastable states) are
readily available. Our approach is the complete opposite of the usual one,
which is to
replace a `dirty' system by an equivalent, `pure' one. We show that the high
temperature (replica symmetric) phase of our `fiduciary' system reproduces
exactly an
approximation due to Golay [\cite{Golay},\cite{Berna}] for the original model,
which,
although unjustified, accounted reasonnably well for Bernasconi's numerical
data at
high enough temperatures [\cite{Berna}]. The entropy given by this
approximation
however becomes negative at low temperatures, signalling, for our fiduciary
model, the
breaking of replica symmetry. We find that the system undergoes a first order
transition towards a low temperature (glass) phase which is rather similar to
the low
temperature phase of Derrida's random energy model [\cite{REM}], although the
entropy
remains non zero - reflecting the fact that small scale motions are not
completely
frozen. This random energy structure is in good agreement with the numerical
findings
of Bernasconi, who found that the energy landscape is `golfcourse' like, with
low
energy states randomly distributed in phase space [\cite{Berna}].
\par We shall first
describe the specific model we considered and its fiduciary disordered version,
and
sketch the main steps of the calculations. We shall then turn to a more
physical (and
speculative) discussion on the relevance of the rather abstract model studied
here for
more realistic situations. \par The model in question
is defined by the following Hamiltonian:
 $$
{\cal H} = {J_0^2 \over 2N} \sum_{k=1}^{N-1} \left [\sum_{i=1}^{N-k} S_i
S_{i+k}\right]^2  \equiv {J_0^2 \over 2N} \sum_{k=1}^{N-1} R_k^2
\eqno(1)$$
where $S_{i=1,.....,N}$ are Ising spins. The scaling of $\cal H$ with $N$ has
been chosen such that $\cal H$ is
extensive.
The spin configurations which minimize  $\cal H$  are binary sequences with
small
autocorrelations, which are useful in communication engineering problems
[\cite{Berna}]. It is difficult to find them because of frustration effects.
\par As mentionned in the introduction, numerical studies show very clearly
that the
system enters a glassy phase at low temperatures, much as if quenched disorder
was
present. Furthermore, the non trivial features of $\cal H$ come from the fact
that the
sum over $k$ extends to infinity when $N \longrightarrow \infty$. In other
words, it
is the couplings between very far away spins which matter - suggesting that the
one
dimensional nature of the problem might not be crucial. We thus propose to
replace Eq.
(1) by the following `fiduciary' Hamiltonian:
$$ {\cal H}_d = {1 \over 2N}
\sum_{k=1}^{N-1} \left [\sum_{i=1}^{N} \sum_{j=1}^{N} J_{ij}^{(k)} S_i
S_{j}\right]^2
\eqno(2) $$
 where $J_{ij}^{(k)}$ are random connectivity matrices, independent for
different $k$'s, with each element equal to $J_0$ with probability $N-k \over
N^2$ and zero
otherwise. This choice insures that the average number of bonds in $\cal H$ and
${\cal
H}_d$ is precisely the same: note that the choice $J_{ij}^{(k)} \equiv J_0
\delta_{i+k,j}$ reproduces exactly Eq. (1). ($J_0$ is set to 1 in the sequel).
${\cal
H}_d$ can be considered as the mean field version of $\cal H$ where the
geometry is lost.
Interestingly, this mean field Hamiltonian allows one to use the replica
formalism. After rather standard manipulations [\cite{MPV}], we find that the
free-energy $F_d$  at temperature $1/\beta$
is given by $- {1 \over N\beta }  \lim_{n \to 0}
{\partial \overline {Z^n} \over
\partial n}$, where the average over the disorder of the n-th power of the
partition function is: $$
\overline {Z^n} = \int \prod_{a < b} dq_{ab} d\hat q_{ab} \left({\rm Tr}_{S_a}
\exp
\sum_{a < b} \hat q_{ab} S_a S_b \right)^N $$
$$ \exp\left[ -N \sum_{a < b}
q_{ab} \hat q_{ab} + N \int_0^1 dx \log \{ \int \prod_a d\lambda_a
e^{-{\beta \over 2}[(1+\beta(1-x))\sum_a \lambda_a^2 + \beta(1-x)\sum_{a
\neq b} q_{ab} \lambda_a \lambda_b ]} \} \right]  \eqno(3)
$$
 Let us first describe  the replica symmetric saddle point
of (3), with $q_{a \neq b}  \equiv q$ and $\hat q_{a \neq b} \equiv \hat q$. We
find
that $q_{\rm saddle} = \hat q_{\rm saddle} = 0$, leading to $F_d^{RS} =
-\beta^{-1}
\int_0^1 dx \log [1 + \beta (1-x)] $. Interestingly, this free energy coincides
{\it
exactly} with the one obtained by Golay [\cite{Golay}] for the original model
Eq. (1),
under the (unjustified) assumption that $R_k \equiv \sum_{i=1}^{N-k} S_i
S_{i+k}$ are
Gaussian independent variables. As shown numerically by Bernasconi, $F_d^{RS}$
gives a
rather good description of the `high' temperature region. This solution however
suffers from the usual entropy disease, which becomes negative below a certain
temperature $T^* = 0.047564...$ and goes to $-\infty$ for $T=0$. However,
 there is no sign of local instability, suggesting  that the
transition to a replica symmetry broken phase must be  first-order (from the
point of
view of the order parameter function: as we shall see, the transition is second
order from the thermodynamical point of view).
 The existence of a phase transition
is ensured by the fact that $\overline{Z^2} \simeq \overline{Z}^2$ at high
enough
$T$.  The one step replica symmetry broken solution allows to introduce, as
usual, a
minimal and a maximal overlap $q_0,q_1$, as well as the position of the
`breakpoint'
$m$, connected with the density of low-lying states (and in turn with the
dynamical
properties [\cite{aging},\cite{CuKu}]). We find that $q_0 \equiv 0$ and $m(T)
\simeq {T
\over T_g}
$ with $T_g = 0.047662 > T^*$, while $y(T) \equiv \beta(1-q_1^2)$ behaves as
shown in Fig 1. Fig 2-a, 2-b show the free energy and the entropy in the low
temperature phase $T < T_g$. Note that the entropy is everywhere positive but
rather
small ($\simeq 10^{-5}$ per spin), goes to zero linearly with $T$ (as in real
glasses), and matches that of $F_d^{RS}$ at $T=T_g$. $T_g$ thus appears as a
freezing
temperature at which $q_1$ discontinuously jumps from zero to a value rather
close to
1 (Note the scale in Fig. 1). The specific heat also jumps at $T_g$. The
picture of
the glass phase is rather similar to that of the random energy model, for which
[\cite{REM},\cite{MG}] $q_0 \equiv 0$, $m(T) \equiv {T \over T_g}$ but also
$q_1
\equiv 1$ corresponding to the fact that the entropy of the frozen phase is
strictly
zero - at variance with our model for which a residual entropy remains. Our
prediction
for the ground state energy is $E_{GS} = 0.02028455...$. Bernasconi noted that
the
numerical ground state energy was much higher than this value as soon as $N
\geq
50$, which might simply reflect the fact that if the energy landscape is that
of the random
energy model, it is extremely difficult to find the ground state.
\par\par It
could be that a more complicated replica symmetry breaking scheme is needed.
We however think that the numerical difference with our
results are likely to be extremely small; it is furthermore irrelevant to the
point addressed
in this letter.  We shall now discuss the above results from a more physical
point of
view. Although quite remote from reality, the model we considered illustrates
the fact
that a pure model can undergo a glass transition which can be described using
the
tools of disordered systems since its mean field formulation naturally
introduces
random variables. Breaking of replica symmetry indicates, as usual, the
existence of
many (metastable) states, and its physical meaning is best understood in the
`cavity'
approach [\cite{MPV}], which is essentially based on a certain (hierarchical)
construction of the equilibrium states and the local field distribution. In
disordered
systems, this method is {\it in fine} equivalent to the replica calculation. In
the
pure model at hand, however, the cavity method in fact allows to recover all
the above
results without introducing a fiduciary random Hamiltonian, but rather through
adequate hypothesis on the statistics of the $S_i$ and the $R_k$ [\cite{us}].
Within a
one pure state picture (`replica symmetric'), the only viable assumption is
that $<S_i>
\equiv 0 $, which immediately leads back to Golay's approximation. The
existence of
many `states' $\alpha$ with weight $P_\alpha$ however allows to go beyond this
result, since it is possible
to have $\sum_\alpha P_\alpha <S_i>_\alpha \equiv 0$ but $\sum_\alpha P_\alpha
<S_i>^2_\alpha = q_1 \neq 0$. In other words, if the spins are frozen in a
given
state, the field acting on the extra (`cavity') spin will be much like a
quenched
random variable. A simplistic way to express this idea might
be the following: one can rewrite Eq.(1) as  $$ {\cal H}  = \sum_{i,j} {J_{ij}
\over 2
\sqrt{N}} S_i S_j \eqno(5) $$ with $J_{ij} \equiv {J_0 \over \sqrt{N}} \sum_k
S_{i+k}S_{j+k}$, i.e. as a spin-glass SK Hamiltonian but for which the
couplings {\it are
themselves determined by the spins}. However, if the dynamics of the system is
slow
(which it is in the `spin-glass' phase) then the couplings can be
self-consistently
thought of as quenched random variables. This suggests that the scenario
described
here is far more general [\cite{Gotze}] and that a genuine short range model of
glass could be a pure
four spin interation Hamiltonian of the form ${\cal H}_4 = - J \sum_{<ijkl>}
S_i S_j
S_k S_l$, where $<ijkl>$ denotes, for example, nearest neighbours tetrahedra on
a
cubic lattice.  ${\cal H}_4$ obviously posseses a `crystalline' ground state
$S_i
\equiv 1$. If, however, the system is quenched from high temperature, the
$\{S_i\}$
are initially random and generate random effective couplings $J_{ij}$, which,
if the
temperature is sufficiently small, will very slowly evolve and lead to a `self
consistent' spin-glass. Numerical simulations [\cite{Sle}] show that this is
indeed
the case: glassy dynamics and aging very similar to that observed in
experimental spin
glasses is clearly observed. Of course, if the interaction is of finite range,
this
quenched effective disorder will progressively anneal out (possibly on
astronomical
time scales), allowing the system to find its crystalline state - as indeed in
real
glasses. Only if the interaction is of infinite range, like in the models
considered
above, or if some topological contraints forbid the annealing of disorder
[\cite{Rivier}] will this glass transition acquire a precise thermodynamical
meaning.
It is clear, however, that the range of interactions need not be very large for
this
limit to be relevant to experimental time scales [\cite{RqK}]: in this work, we
have
primarily focused on static calculations, leaving the investigation of the
dynamics
(along the lines of [\cite{FM},\cite{CuKu0}]) for future work. A dynamical
approach
is clearly needed in order to identify which internal degrees of freedom get
quenched and to make a link with the mode coupling theory [\cite{Gotze}].
The success of the present approach might indicate that the
precise decomposition of
which degrees of freedom become  quenched or annealed is maybe not so crucial.

\vskip 2cm
Acknowledgments.
We are grateful to G. Parisi who pointed out to us long ago that this model
has a glassy behaviour.
We thank D. Dean, W. Krauth, S. Slijepcevic and E. Vincent for helpful
discussions.
While completing this work, we learnt that our colleagues and friends in Rome,
 E. Marinari,  G. Parisi and F. Ritort were developing rather similar
ideas on a version of this model with periodic boundary conditions
[\cite{mapari}].
MM thanks the SPhT of the CEA Saclay for its hospitality.
 \vskip 2cm Figure captions.\par Fig 1: Plot of the quantity $y(T) \equiv {(1-
q_1^2(T))\over T}$ in the low temperature phase. Note the scale, which
indicates that the deviations of $q_1$ from 1 are very small. \par Fig 2: Plot
of the free-energy
(2-a) and entropy (2-b) in the glass phase. Note that $S(T)$ is positive but
rather
small. We found that $S(T) = A T$  (for small $T$) with $A \simeq 5 \times
10^{-5} $\par
\references

{
\refis{Houches} `Liquids, freezing and glass transition', Les Houches 1989, JP
Hansen, D. Levesque, J. Zinn-Justin Editors,  North
Holland, in particular ref. [\cite{Gotze}]
\refis{AHV} see e.g. P. W. Anderson, in Les Houches 1979, R. Balian, R. Maynard
and G. Toulouse Editors, North Holland.
\refis{Thiru} T. R. Kirkpatrick, D. Thirumalai, P. G. Wolynes, Phys. Rev. A 40,
1045
(1989) and references therein.
\refis{MPV} M. M\'ezard, G.
Parisi, M.A. Virasoro, {\it Spin-glass theory and beyond}, World Scientific,
1987 (Singapore).
\refis{Orland} T. Garel, H. Orland, Europhys. Lett. 6, 307 (1988), D.
Thirumalai,
T. Garel, H. Orland, in preparation.
\refis{Shak} E. I. Shakhnovich, A. M. Gutin, Europhys. Lett. 8, 327  (1989)
\refis{Wolynes} see e.g. H. Frauenfelder, P. Wolynes, Physics Today,
p. 58 (February 1994) and references therein.
\refis{Goldbart} P. Goldbart, N. Goldenfeld, Phys. Rev. Lett. 58 (1987)2676;
Phys.
Rev. A39 (1989) 1402.
\refis{Gotze} W. Gotze, in ref. [1], p. 403 et sq., p. 458. The mode coupling
theory attemps to capture `self-blocking' effects, quite similar to the self
induced disorder discussed here.

\refis{Golay}  M.J.E. Golay, IEEE IT-23 (1977) 43, IEEE IT-28 (1982) 543.
\refis{Berna} J. Bernasconi, J. Physique 48 (1987) 559.
\refis{MW}  W. Krauth, M. M\'ezard, in preparation.
\refis{REM} B. Derrida, Phys. Rev. B 24 (1981) 2613
\refis{aging} J. P. Bouchaud, J. Physique I 2  (1992) 1705
\refis{CuKu} L. Cugliandolo, J. Kurchan, Phys. Rev. Lett. 71 (1993) 173
\refis{MG} D. Gross, M. M\'ezard, Nucl. Phys. B 240 (1984) 431
\refis{us} J.P. Bouchaud, M. M\'ezard, unpublished.
\refis{Sle} S. Slijepcevic, J. P. Bouchaud, in preparation.
\refis{Rivier} N. Rivier, Adv. Phys.  36 (1987)96
\refis{RqK} It is interesting to note that if the sum over $k$ extends up to
$K$ in
Eq. (1), the energy to create a domain wall between two favourable
configurations
grows proportionaly to $K$; the  time scale for jumping from one
low lying configuration to another one by sweeping a wall thus diverges as
$e^K$.
\refis{FM} S. Franz, M. M\'ezard, preprints ENS 93-39, 94-05.
\refis{CuKu0} L. Cugliandolo, J. Kurchan, Rome University preprint 977 .
\refis{mapari} E. Marinari, G. Parisi, F. Ritort, in preparation.

}
\endreferences
\end